# Emamectin benzoate sensing using vivianenes (2D vivianites)

*Surbhi Slathia[a], Bruno Ipaves[b], Raphael Benjamim de Oliveira[b], Guilherme da Silva Lopes Fabris[b], Marcelo Lopes Pereira Júnior[c], Raphael Matozo Tromer[b], Gelu Costin[d], Suman Sarkar[e], Douglas Soares Galvão[b*], Chandra Sekhar Tiwary[f, a*]*

[a]School of Nanoscience and Technology, Indian Institute of Technology Kharagpur, Kharagpur 721302, India.

[b]Applied Physics Department and Center for Computational Engineering & Sciences, State University of Campinas, Campinas, Sao Paulo 13083-970, Brazil

[c]University of Brasília, College of Technology, Department of Electrical Engineering, Brasilia, 70910900, Brazil.

[d]Department of Earth, Environmental and Planetary Sciences, Rice University, Houston, TX 77005, United States of America

[e]Department of Materials Engineering, Indian Institute of Technology Jammu, Jammu 181221, India.

[f]Department of Metallurgical and Materials Engineering, Indian Institute of Technology Kharagpur, Kharagpur, West Bengal 721302, India.

**Abstract**

The excessive application of pesticides, particularly the overreliance on insecticides for the protection of desirable crops from pests, has posed a significant threat to both ecological systems and human health due to environmental pollution. This research outlines a comprehensive approach to recognizing and quantifying the presence of insecticides through the application of spectroscopic and electrochemical sensing methods. The detection of Emamectin benzoate (EB), a commonly used insecticide, was performed utilizing vivianenes, a 2D phosphate that has been mechanically exfoliated from the naturally occurring vivianite minerals. This investigation examined the structural and compositional characteristics of vivianenes, utilizing a range of characterization methods. The spectroscopic analyses reveal the molecular interactions and structural modifications that take place during the interaction of EB with the 2D template. Electrochemical investigations employing cyclic voltammetry were performed for different concentrations of EB to enable real-time monitoring of the pesticide. The modified sensing electrode using vivianene demonstrated a linear range of from 50 mg/L to 10 μg/L, effectively detecting EB molecules at levels significantly below the hazardous threshold. Fully atomistic molecular dynamics simulations were also carried out to obtain further insights into the interaction mechanisms of the EB with the vivianites, and the results corroborate the adsorption mechanism. Our results highlight the potential application of 2D phosphate minerals as advanced sensors to enhance agricultural monitoring and promote sustainable development.

**Introduction**

As public exposure to biological, chemical, and physical threats due to industrial globalization and population growth is increasing day by day, the emergence of nano-sensors is paving the way for a new era of ultrasensitive and real-time monitoring. The identification and surveillance of radiation, hazardous chemicals, pathogens, and pesticides present a significant challenge on a global scale [1,2]. The utilization of two-dimensional (2D) materials in sensor technology is experiencing significant growth, spanning applications from environmental monitoring to healthcare, attributed to their distinctive properties derived from their fundamental structures [3,4]. 2D materials are characterized by their layered structure, typically measuring just a few nanometers in thickness, resulting in an exceptionally high surface-area-to-volume ratio. This leads to the availability of multiple reactive sites between the material and analytes. Moreover, the conductivity of 2D materials can be tuned by manipulating structural defects, varying the number of layers, introducing dopants, or applying post-functionalization techniques to the material [5,6]. More importantly, the mechanical strength [7,8] and flexibility of 2D nanomaterials render them suitable for integration with advanced technologies, including ultrathin silicon channels, various printing techniques, metal electrodes, and flexible as well as wearable electronics [9–12].

The mechanical exfoliation of graphite to isolate single layers of graphene has been accompanied by significant experimental endeavors aimed at exfoliating other layered bulk crystals [13]. This pursuit seeks to create a diverse array of 2D materials that possess complementary properties to those of graphene. The investigation of additional 2D materials is driven not only by the necessity to identify substances with properties tailored for specific applications but also by the potential to uncover intriguing and occasionally unforeseen physical phenomena during these exploratory surveys .

Phosphates are minerals characterized by the presence of the tetrahedrally coordinated phosphate ($PO_4^{3-}$) anions. The vivianite group of minerals holds significance for several reasons, which include their widespread occurrence in nature, in soils and sediments, and their photosensitivity and photo-self oxidation [14]. Vivianites are hydrated iron phosphates characterized by the approximate formula $Fe^{2+}_3(PO_4)_2.8H_2O$. Vivianites exhibit sensitivity to exposure to visible light, resulting in a significant alteration in their color, from colorless or pale green to dark green or brown. Bulk vivianites serve as a natural electron donor, demonstrating effectiveness in the dechlorination of various chlorinated organics, which are the primary and most commonly encountered contaminants in soil and groundwater, contributing to considerable environmental challenges [15,16]. However, there is a lack of experimental reports regarding their thin flakes in the existing literature. The investigation of their properties upon mechanical exfoliation would be intriguing, particularly given the layered structure observed in the bulk form.

Among the numerous environmental issues currently confronting us, the exposure to pesticides presents a significant challenge. Pesticides play a vital role in managing agricultural pests and securing crop production; however, they can remain in ecosystems, infiltrating water sources and accumulating in soil and the food chain. Prolonged exposure to pesticides, even at low concentrations, presents significant health risks to humans. These risks encompass neurological, respiratory, and other disorders, along with an elevated likelihood of cancer. In addition, the application of pesticides has significant implications for biodiversity, adversely impacting non-target species, including beneficial insects, aquatic organisms, and avian populations, which can result in ecological imbalances. The increasing dependence on agricultural output necessitates the establishment of safe pesticide residue levels, which is vital for safeguarding public health and maintaining environmental integrity.

Emamectin benzoate ((4"R)-4"-deoxy-4"-(methylamino) avermectin B1 benzoate) (EB) is a macrocyclic lactone insecticide, specifically designed for the management of lepidopteran insects and various other pests [17,18]. EB exhibits significant persistence in the sediments of aquatic systems, attributed to its extended degradation half-life and hydrophobic characteristics. This compound poses toxicological risks to aquatic organisms, resulting in enduring effects on their health and ecosystems. The impact on soil fertility is detrimental, as it inhibits the metabolic activities of microorganisms that inhabit the soil. It has been observed that the introduction of EB into soil renders the soil microbes unable to utilize the organic matter present in the soil [19–21].

In this study, we show that we can detect EB utilizing vivianenes (2D vivianites) flakes that were mechanically exfoliated from bulk vivianite mineral through the bath sonication method. The exfoliated samples underwent characterization through X-ray diffraction (XRD), X-ray photoelectron spectroscopy (XPS), and Raman spectroscopy to assess their structural properties, elemental composition, and vibrational modes. The investigation into the morphology and layer arrangement was conducted utilizing scanning electron microscopy (SEM) and transmission electron microscopy (TEM). A dual-mode analysis was conducted for pesticide detection, incorporating both spectroscopy and electrochemical methods. The various concentrations of EB were analyzed to determine the detection limit through Raman and UV-VIS spectroscopies. The binding mechanism was assessed through FTIR analysis. The electrochemical sensing was conducted using cyclic voltammetry with a three-electrode setup comprising a counter, reference, and working electrode. Classical molecular dynamics (MD) simulations were performed to gain further insights into the interaction mechanism between EB molecules and the surface of vivianene flakes.

**Materials and Methods**

The vivianite crystals were collected from the Tip Top Mine located in South Dakota, USA. The bulk crystals were first ground into a fine powder using a mortar and pestle. 30 mg of the bulk powder was introduced into 30 ml of isopropyl alcohol (IPA, also known as 2-propanol, Merck >99% purity, M = 60.10 g mol$^{-1}$). The solution underwent immersion in a bath sonicator for approximately 5 hours. The temperature was consistently kept below 35 °C. After sonication, the solution was allowed to remain undisturbed for 24 hours to facilitate the settling of the precipitates. The supernatant, which is the liquid portion, was gathered for subsequent analyses and experimentation.

The slurry utilized in the electrochemical experiments was prepared by combining the supernatant solution with PVDF in a 4:1 ratio, where PVDF acted as a binder. The slurry underwent sonication for 30 minutes to achieve comprehensive mixing. A small portion of this slurry was applied to a bare GCE electrode through the drop casting method and subsequently dried for 2 hours for the electrochemical detection of Emamectin Benzoate (EB).

The crystallographic structure and phase identification of both bulk and 2D phosphoferrite were determined using an X-ray diffractometer (Bruker D8 Advance) with Cu Kα radiation, featuring a wavelength of λ = 0.15406 across a 2θ range of 15–55°. The atomic arrangements were examined utilizing a high-resolution transmission electron microscope (HRTEM) operating at 230 kV. Ambient temperature investigation of Raman spectra was conducted utilizing a WiTec UHTS Raman spectrometer 600 VIS, Germany, configured with an excitation wavelength of 532 nm. The standard absorption spectra were obtained utilizing an analytical UV–visible (UV–vis) spectrophotometer. The surface morphology of both the bulk and 2D samples was examined utilizing scanning probe microscopy (FEI Quanta 400), while elemental mapping was conducted through energy dispersive x-ray spectroscopy (EDS). The interactions between the exfoliated sheets and EB molecules were examined utilizing a Thermo Fisher Scientific Instruments FT-IR spectrometer. Cyclic voltammetry measurements were conducted using Ag/AgCl as the reference electrode, platinum as the counter electrode, and a coating of (Vivianene + PVDF) on a glassy carbon electrode (GCE) using the Squidstat Plus user interface (potentiostat/galvanostat) in a three–electrode cell configuration.

**Modeling methodology**

To investigate the interaction of the EB with 2D vivianites, we have used classical molecular dynamics simulations using the Forcite software, as implemented in the Materials Studio suite [22]. For the structural optimization, it was used the steepest descent algorithm [23], with convergence criteria for energy, force, stress, and displacement set at $2.0\times10^{-5}$ kcal/mol, $1\times10^{-3}$ kcal/mol/Å, $1\times10^{-3}$ GPa, and $1.0\times10^{-5}$ Å, respectively.

To model the atomic interactions of the system, the universal force field (UFF) [24] was employed, along with the QEq method, to calculate atomic charges [25] in the equilibration process, considering a convergence limit of $1.0\times10^{-4}$ e$^-$. The non-bonded interactions were described by considering electrostatic interactions using the Ewald method [26,27], with an accuracy of $1.0\times10^{-5}$ kcal/mol and a buffer width of 0.5 Å. van der Waals (vdW) interactions were also taken into account, using the atom-based summation method and a cubic spline truncation approach [28], with a cutoff distance of 18.5 Å, and spline and buffer widths of 1 Å and 0.5 Å, respectively.

In addition to optimizing the vivianite crystal, EB (see **Figure S1b**) was also optimized using the parameters/protocols previously described. The molecular system contains 152 atoms and has the chemical formula $C_{56}H_{81}NO_{15}$.

To investigate the adsorption of the insecticide on the different possible faces of the vivianite crystal, the crystal was cleavaged into three distinct two-dimensional systems: the xy, xz, and yz planes. In this process, symmetric supercells with dimensions of 3 × 3 × 2, 3 × 1 × 7, and 1 × 3 × 7 were created (see **Figure S1a**). A vacuum buffer layer of 50 Å was introduced to avoid spurious interaction with mirror cells.

**Results and Discussion**

The chemical formula of vivianites is often represented as $Fe_3(PO_4)_2 \cdot 8H_2O$. $Fe^{2+}$ occupies two distinct locations within the crystal structure of vivianite. At one location, $Fe^{2+}$ ions are coordinated by two trans-oxygen ligands from $[PO_4]^{3-}$ tetrahedra and four $H_2O$ ligands, resulting in a singular $Fe_a$ octahedron [16]. At an alternative location, $Fe^{2+}$ coordinates with two $H_2O$ ligands and four oxygen atoms from the $[PO_4]^{3-}$ tetrahedron to create a $Fe_b$ octahedron, with a common oxygen-oxygen edge between each pair of $Fe_b$ octahedrons as shown in **Figure 1a** [30,31]. The iron in the structure can be substituted with trace amounts of various metal elements, including manganese, magnesium, zinc, nickel, and calcium. The vivianite structures consist of chains of $FeO_6$ octahedra linked by $PO_4$ tetrahedra, with interlayer water molecules facilitating separation and hydrogen bonding among the layers. The structural composition of vivianites consists of interconnected chains of $FeO_6$ octahedra, which are linked by $PO_4$ tetrahedra, along with interlayer water molecules that contribute to its overall architecture. The structural integrity of vivianites is maintained through comparatively weak van der Waals interactions and hydrogen bonding that occur among the water molecules and the oxygen atoms associated with the phosphate groups. This arrangement allows the layers to be separated with relative ease, particularly when solvents or mechanical forces are applied during the exfoliation process.

The SEM image of the bulk crystal is presented in **Figure 1c,** accompanied by an adjacent image in **Figure 1d** confirming the layered structure of the vivianite mineral. **Figure 1e** illustrates the exfoliated vivianene, exhibiting a lateral dimension of approximately 0.25 to 0.4 µm. The EDX calculations for bulk vivianite and exfoliated vivianene were performed as shown in SI **Figure S2**. Traces of manganese were also detected in the case of bulk, and the concentration decreased in the exfoliated sample.

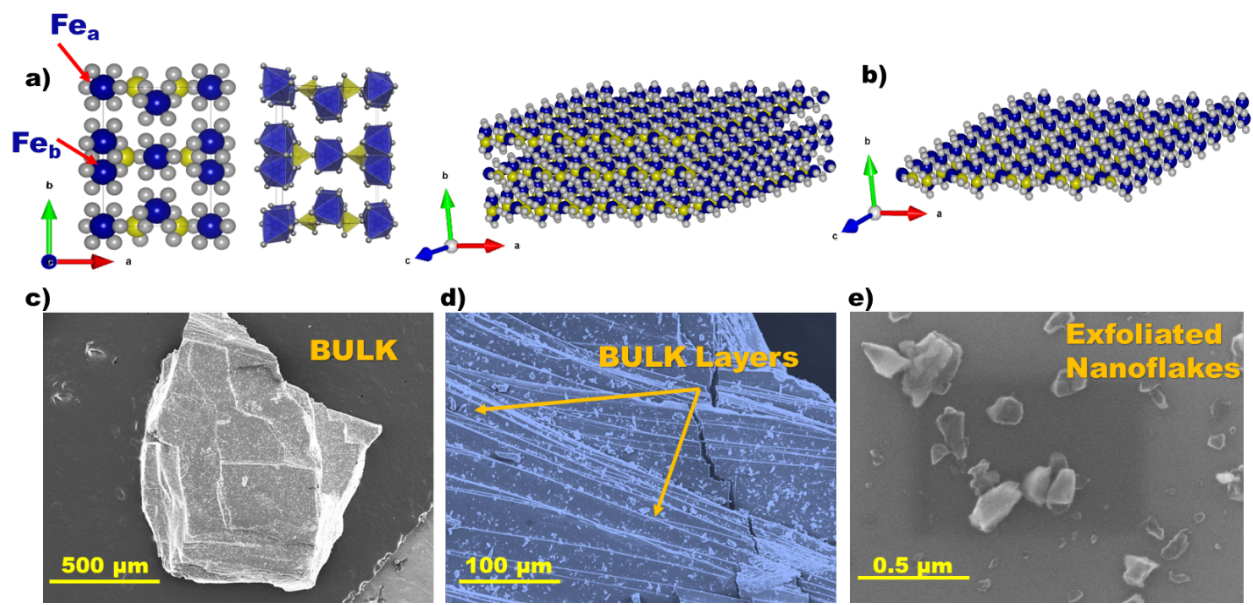

**Figure 1**: Physical morphology of bulk and two-dimensional vivianites. a) Diagram depicting the synthesis process of 2D nanoflakes from bulk material, b) Schematic of a single layer of vivianene, c) SEM image of the bulk sample along with d) highlighting its layered structure, and e) SEM image showcasing the exfoliated sheets of vivianites.

In **Figure 2a**, we present the XRD analyses of both the bulk and exfoliated vivianite ore within a 2θ range of 15º to 60º. The monoclinic structure, characterized by lattice parameters a = 10.086 Å, b = 13.441 Å, and c = 4.703 Å, and belonging to the space group C12/m1, is seen in bulk form. Significant peaks were detected at 39.07º and 54.26º, which matched the (420) and (441) planes in the powdered bulk sample. Following the exfoliation of the bulk sample, a notable decrease in the intensity of the majority of diffraction planes was observed, suggesting a decrease in interlayer interactions and leading to a shift from a three-dimensional bulk structure to a two-dimensional morphology. The absence of these planes suggests that the exfoliated material is predominantly composed of a few layers. The pronounced peaks at 38.39º and 44.59º were observed in the exfoliated sample, signifying that the predominant nanosheets were oriented along the (311) and (260) crystallographic directions, respectively.

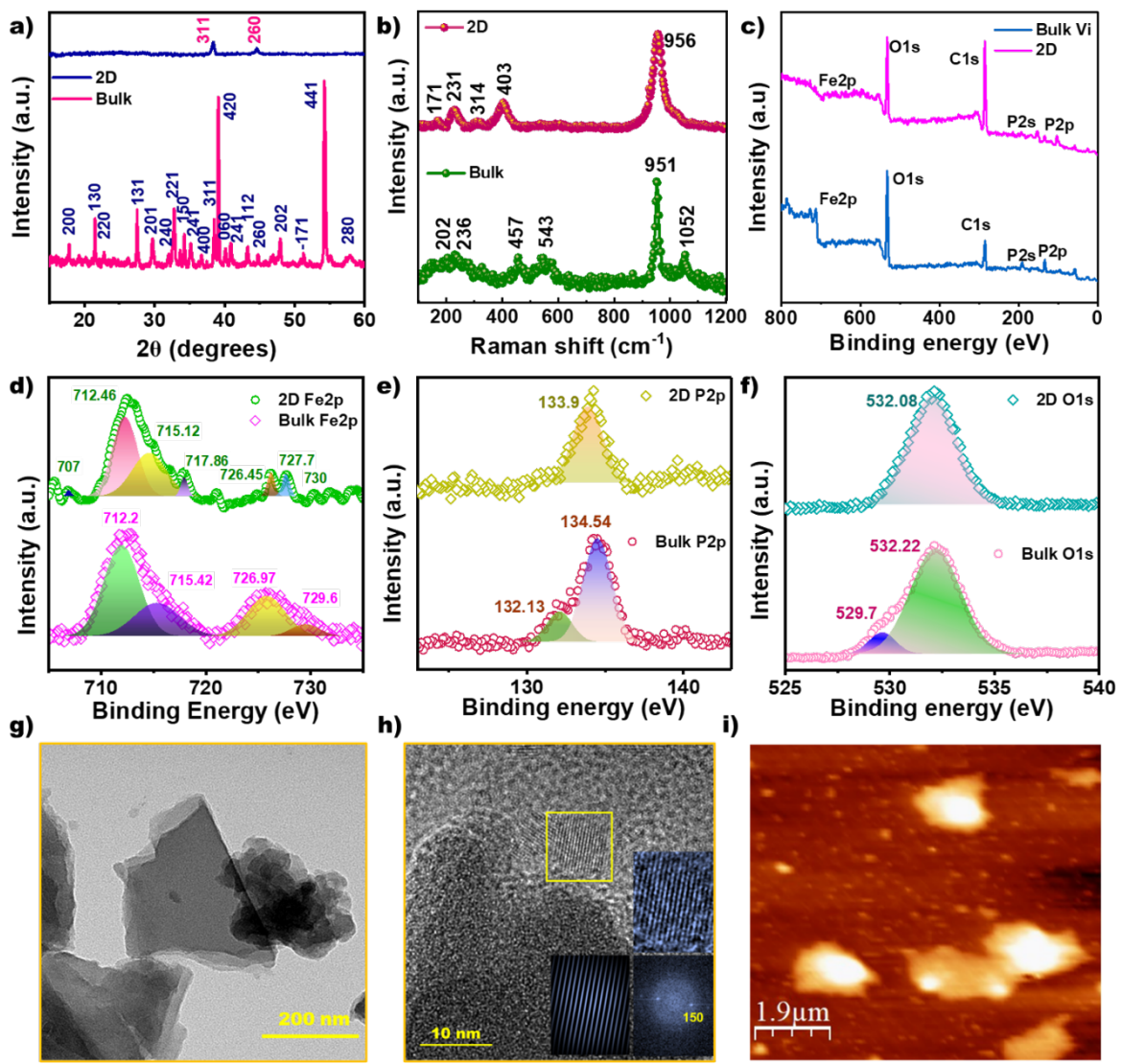

**Figure 2:** Structural and compositional analyses of vivianite and vivianene. a) XRD pattern of bulk and 2D sample, b) Raman spectrum of the bulk vivianite and vivianene. c) XPS investigations of both bulk and 2D samples, along with the deconvoluted spectra from **Figure**

**(d-f)** of Fe 2p, P 2p, and O 1s, respectively, g) HRTEM image of the exfoliated sheets, h) The FFT and inverse FFT analyses of the planes on the surface of vivianene, and i) AFM image of the vivianene flakes with inset showing an average thickness of layers, approximately 8.5 nm.

In order to investigate the vibrational modes present in both bulk and 2D vivianites, we conducted Raman spectroscopy analyses, as illustrated in **Figure 2b**. The Raman analysis of the bulk sample revealed a strong band at 951 cm$^{-1}$, accompanied by a weaker one at 1052 cm$^{-1}$ within the phosphate region. The band at 951 cm$^{-1}$ is attributed to the Raman active PO-stretching ($\upsilon_1$) vibration in the A$_g$ band. The location of this band aligns remarkably well with the results published earlier [32]. The spectral band observed at 1052 cm$^{-1}$ is attributed to the antisymmetric ($\upsilon_3$) stretching vibrations of the phosphate PO group. It is important to highlight that in the case of 2D, a single peak with a slight shift at 956 cm$^{-1}$ was observed, while the low-intensity peak vanished. Additionally, the broadened peak at 956 cm$^{-1}$ includes a small peak at 1022 cm$^{-1}$, which corresponds to the B$_g$ mode of antisymmetric ($\upsilon_3$) PO vibrations. The observed low-intensity peak at 543 cm$^{-1}$ is associated with the bending modes ($\upsilon_4$) of the PO$_4$ and/or Fe/H$_2$O, and the peak at 457 cm$^{-1}$ is the $\upsilon_1$ bending of PO$_4$ [14]. In the Raman spectrum for vivianite, low-intensity bands are observed between 170 cm$^{-1}$ and 250 cm$^{-1}$, which may be attributable to FeO-stretching vibrations.

The elemental analysis and the oxidation states present in the samples were assessed using XPS, as displayed in **Figures 2c** to **2f**. The mineral is primarily composed of iron, phosphorus, and oxygen, exhibiting distinct peaks at approximately 712 eV (Fe 2p), 532 eV (O 1s), and 134 eV (P 2p) for the bulk and 2D samples, respectively. The high-resolution XPS spectra of Fe 2p are presented in **Figure 2d**. For the bulk sample, the Fe 2p$_{3/2}$ and Fe 2p$_{1/2}$ doublet peaks were observed at 712.2 eV and 726.97 eV, respectively. The satellite peaks corresponding to Fe$^{2+}$ appeared at approximately 715.42 eV and 729.6 eV [33–35]. The peak observed at 712 eV is indicative of the presence of FePO$_4$, consistent with findings from prior investigations [36]. The spectrum of P 2p (**Fig. 2e**) reveals peaks at 132.13 and 134 eV, which are associated with $PO_4^{3-}$ and $HPO_4^{2-}$, respectively. The spectra of P 2p (**Figure 2e**) reveal peaks at 133.9 eV (2D) and 134.52 eV (bulk), which are associated with $PO_4^{3-}$. The small peak observed around 132.12 eV may have resulted from organic phosphorus, primarily released from deceased microbial cells and phosphorus-accumulating organisms in the context of natural minerals [37]. The peak in the O 1s spectra at 532.22 eV and 529.7 eV corroborates the presence of P-O and Fe-O bonds, respectively. In the case of 2D, a singular significant peak for the P-O bond was observed in **Figure 2f** [38].

**Figure 2g** presents the high-resolution TEM image of the exfoliated sheets derived from bulk vivianite, exhibiting several layers. The identification of the planes on the sheets was achieved through the application of FFT and inverse FFT techniques on the magnified image of a flake. The d-spacing of the plane was determined to be approximately 0.21 nm, with the plane orientation identified as (150).

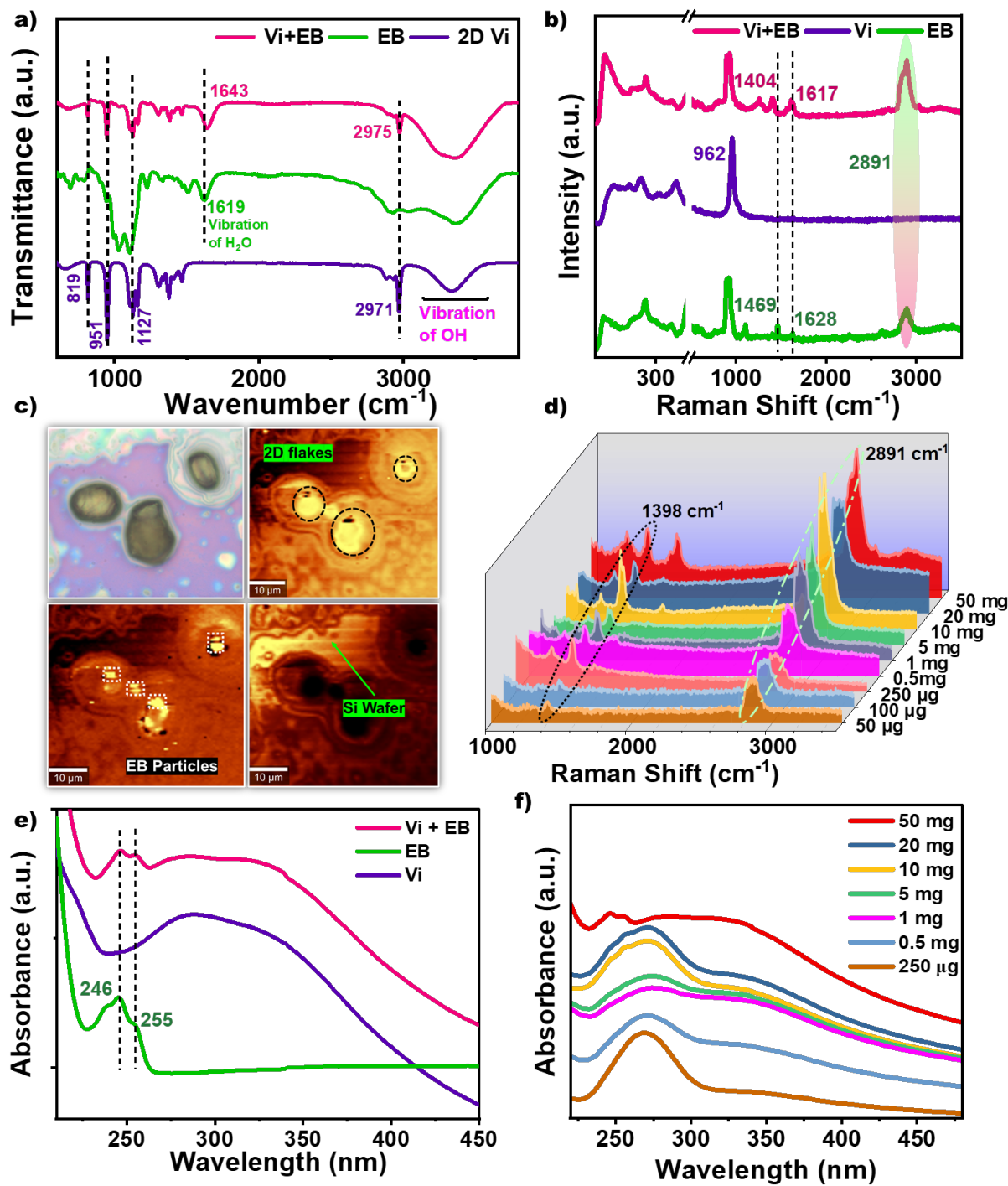

**Figure 3:** Spectroscopic analyses of EB mixed with vivianene. a) FTIR spectra of vivianene, EB, and the mixture of vivianene with EB; b) Raman spectra of vivianene, EB, and the mixture of vivianene with EB drop-cast on a silicon wafer; c) Raman mapping to analyze the presence of individual samples; d) Raman spectra of varying concentrations of EB ranging from 50 mg/L to 50 μg/L; e) UV-VIS spectra of vivianene, EB, and the combination of vivianene and EB; f) UV-VIS spectra obtained for different concentrations of EB.

In **Figure 3a**, we present the FTIR spectra for vivianene, EB, and the combination of vivianene and EB. The broad O-H stretching of the Fe-OH group in vivianene was observed in the range of 3581 to 2881 cm$^{-1}$. The observed peaks at 1158, 1127, and 951 cm$^{-1}$ in the phosphate region are attributed to the $\upsilon_3$ and $\upsilon_1$ stretching modes of $PO_4^{-3}$. The peak observed at 815 cm$^{-1}$ is attributed to the libration of OH groups [39,40]. In the case of EB, the broad peak observed between 3705 and 3131 cm$^{-1}$ is attributed to O-H stretching vibrations, while a smaller peak at 2920 cm$^{-1}$ is associated with C-H stretching vibrations from the aromatic ring of the benzoate portion. The peak observed at 1619 cm$^{-1}$ is attributed to the bending vibrations of absorbed water molecules. The peak observed at 1506 cm$^{-1}$ corresponds to the carbonyl group (C=O), while the peaks at 1413 and 1331 cm$^{-1}$ are attributed to the asymmetrical and symmetrical bending vibrations of C-H, respectively. The peaks observed at 1104 and 1029 cm$^{-1}$ are attributed to O–H and C–O–C flexion, while the peak ranging from 947 to 510 cm$^{-1}$ is associated with C–H flexion occurring outside the plane in an aromatic ring or a C=C cis bond [41–44]. It is important to note that when EB was added to vivianene, no new peaks were observed, indicating that most of the interaction between the molecules is likely due to physical adsorption only. Spectroscopic techniques in sensing provide a fast, non-contact, and non-destructive approach to detection. The distinct fingerprint spectra of an analyte facilitate the identification of specific compounds through their vibrational and electronic transitions. Raman spectroscopy proves to be a valuable technique for identifying analytes by observing the alterations in various vibrational modes of functional groups present in the target molecules. **Figure 3b** shows the spectra of EB, vivianene, and vivianite + EB drop cast on a silicon wafer. The peak observed at 2891 cm$^{-1}$ in the case of EB may be attributed to the stretching mode of C-H vibrations within the benzene ring. The in-plane C=C stretching has resulted in the peak at 1628 cm$^{-1}$, while the prominent peaks near 1457 cm$^{-1}$ can primarily be ascribed to in-plane ring stretching vibration bands or $CH_3$ vibrations

[45,46]. The additional peaks observed around 1109 and 920 cm$^{-1}$ are attributed to the C-H bending motion and the coupling of the ring breathing mode with the COO- stretching vibration, respectively [45]. A significant change was observed in two pronounced peaks at 1457 cm and 1628 cm$^{-1}$, which shifted to 1398 and 1617 cm$^{-1}$, respectively. Therefore, we can take into account these results as the fingerprint region for the identification of EB pesticide. **Figure 3c** illustrates the Raman mapping of vivianene with the analyte on a silicon wafer. Observations indicate that EB molecules tend to encompass the vivianene flakes, shedding light on the physical adsorption of EB on the vivianene surface, as corroborated by FTIR analysis. Raman spectra were obtained for various concentrations of EB ranging from 50 mg/L to 50 µg/L, as illustrated in **Figure 3d**. The peak observed at 1398 cm$^{-1}$ exhibited consistency across all concentrations, accompanied by a linear decrease in intensity. Nonetheless, the peak observed at 1617 cm$^{-1}$ disappeared at lower concentrations, potentially attributed to the reduced Raman scattering signal at these concentrations or a decrease in molecular interactions. The UV-VIS spectra of vivianene exhibit a broad peak ranging from 264 to 355 nm, while EB displayed three notable peaks at 238, 246, and 255 nm. Upon the addition of EB to vivianene, the significant peaks of EB became evident, as illustrated in **Figure 3e**. As the concentration of EB was reduced, the peaks exhibited a shift towards a higher wavelength (red shift), approximately 269 nm, as illustrated in **Figure 3f**.

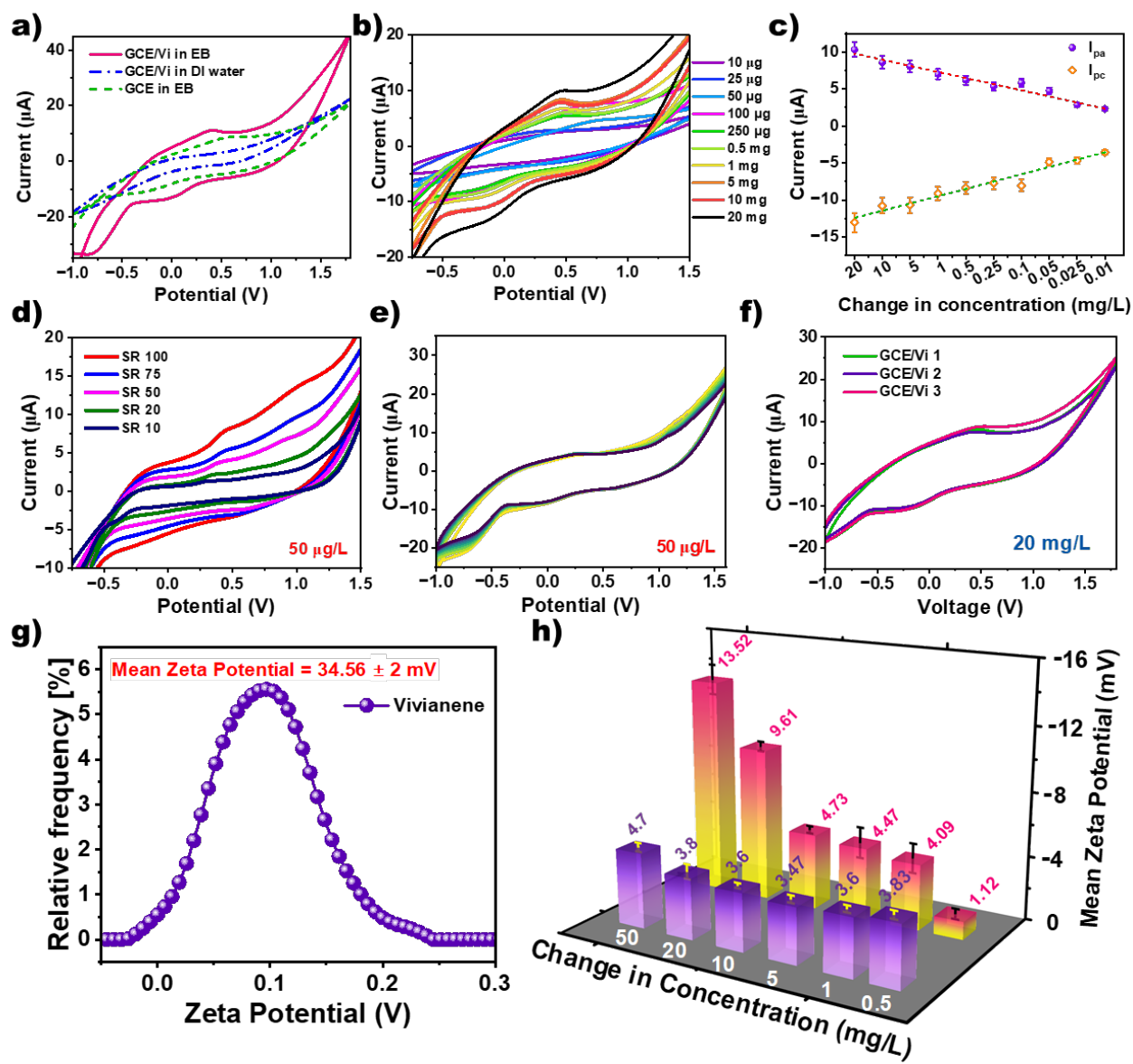

**Figure 4:** Cyclic voltammetry measurements and Zeta potential. a) A comparison of the CV measurements performed with GCE and modified vivianene/GC electrodes in DI water and EB at a concentration of 50 mg/L and a scan rate of 100 mV/s was carried out, b) The CV measurements that were made for different concentrations of EB ranging from 20 mg/L to 10 μg/L; c) Change in oxidation and reduction current as the concentration is varied, d) Variable scan rate measurements for a concentration of 50 μg/L, e) 10 CV cycles were obtained for the vivianene/GCE at a 20 mV/s scan rate in order to investigate the stability for 50 μg/L concentration, f) CV results of different vivianene/GCE electrodes to check the reproducibility, g) Zeta potential of vivianene, and h) change observed in zeta potential for different concentrations of EB with and without vivianene.

Electrochemical sensing has gained prominence over conventional analytical techniques due to its enhanced sensitivity, specificity, and rapid response times. We have performed cyclic voltammetry (CV) experiments using a modified glassy carbon electrode (GCE) coated with 2D vivianite (vivianene) and polyvinylidene fluoride (PVDF) mixture across different EB concentrations. **Fig. 4a** illustrates the cyclic voltammetry response of the modified Vi/GCE in a 50 mg/L solution at a scan rate of 100 mV/s, compared to the vivianene/GCE in DI water and the bare GCE in the EB solution. No oxidation or reduction peaks were detected for vivianene/GCE in deionized water; however, the addition of EB to the solution resulted in the observation of an oxidation peak at a potential of 0.41 V and a reduction peak at -0.1 V. However, bare GCE exhibited a diminished response in the EB solution. The concentration of EB was altered between 10 µg/L and 20 mg/L, and CV measurements were conducted using a modified vivianene/GCE. The results indicated a linear increase in both the oxidation and reduction peak currents as the concentration of EB increased, as illustrated in **Figure 4b**. Observations indicated that at the lower concentrations of 10 µg/L and 20 µg/L, no notable peaks were detected during a scan rate of 100 mV/s. As the scan rate was reduced to 20 mV/s (SI **Figure S2**), the lower concentrations exhibited a response; however, the oxidation peak for the 10 and 5 µg/L samples shifted to a higher potential value of 1.05 V. The relationship between the anodic and cathodic peak current response and the variation in concentration is shown in **Figure 4c**. The fitted linear curve of the plot, exhibiting a regression coefficient value of 0.96 for anodic peak current and 0.95 for cathodic peak current, indicates a direct relationship between the current response and the variation in concentration. Measurements of scan rates are crucial for determining whether the underlying process is governed by diffusion or adsorption mechanisms. Consequently, we conducted measurements by adjusting the scan rate from 100 mV/s to 10 mV/s, as illustrated in **Figure 4d**. To determine if the oxidation and reduction process of EB is diffusion-controlled, a graph of the square root of the scan rate versus the anodic peak current I$_p$ was constructed (**SI Figure S3b**). The peak current exhibited a linear increase with the rise in scan rate, demonstrating a regression coefficient of 0.977, which supports the conclusion that the process is governed by diffusion control. The electrode's stability was assessed through 10 cycles of measurements performed at a scan rate of 20 mV/s, utilizing a concentration of 50 µg/L (**Figure 4e**). The cyclic voltammetry graphs exhibited a consistent response across all cycles, thereby validating the cyclic stability observed in the study. The consistency of the response across multiple measurements was evaluated by preparing three electrodes with identical parameters for comparison, as shown in **Figure 4f**. The observed minor variation in the graph could be attributed to human error; however, the oxidation and reduction peaks consistently appeared at nearly identical potentials across all electrodes.

Zeta potential serves as a crucial metric for elucidating the electrostatic interactions occurring within a sample in a solution. This method offers insights into the net surface charge of the analyte and the analyte sample when they are combined in a solution. The binding of an analyte to a functionalized particle possessing a defined charge allows for the monitoring of shifts in zeta potential. These shifts serve as indicators of successful binding or complex formation, thereby facilitating the recognition and quantification of the analyte. **Figure 4g** illustrates the zeta potential graph pertaining to vivianene. The mean zeta potential of vivianene was determined to be 34 mV, signifying a net positive charge on the surface. The net charge of EB at a maximum concentration of 50 mg/L was observed to be negative, measuring approximately -13.52 mV. The addition of EB to the vivianene solution resulted in a decrease in the net surface charge to -4.7 mV (see **Figure 4h**), suggesting that immediate flocculation is taking place, indicating an interaction between the EB particles and the vivianene sheets. The positively charged vivianene appears to be partially mitigating the negative charges of EB, potentially through mechanisms such as electrostatic attraction or adsorption.

**Simulations of the interaction between vivianene and the pesticide emamectin benzoate**

We have carried out a series of classical molecular dynamics (MD) simulations to investigate the EB interactions with vivianene.

The vivianite crystals obtained based on the previously mentioned protocols and parameters, presenting a three-dimensional monoclinic-B lattice with a C2/m (C2H-3) symmetry group, which is composed of 74 atoms ($H_{32}O_{32}P_4Fe_6$) in the unit cell. The resulting optimized UFF lattice parameters are a, b, and c equal to 9.76 Å, 13.83 Å, and 4.67 Å, respectively, with angles α, β, and γ of 90°, 107.81°, and 90°, respectively. The crystal density is 2.77 g/cm³, with a total energy of -50 kcal/mol/Å. The dimensions of the resulting 2D supercells were 29.28 × 41.48 Å², 29.28 × 32.73 Å², and 41.48 × 32.73 Å² for the xy, xz, and yz planes, respectively. **Figure S1 (a-f)** displays the front and side views of the two-dimensional systems obtained from the xy, xz, and yz planes, respectively.

To understand how the EB adsorption occurs in each of the cleavaged vivianite structures, we have used the adsorption locator module of the Materials Studio suite [47], which consists of scanning the different surfaces of the nanomaterial and calculating the adsorption energy for several configurations [48]. Thus, these configurations are evaluated through Monte Carlo simulations of the various substrate-adsorbate combinations as the temperature varies. Within each temperature range, the charge and other components of the adsorbate are kept fixed. This type of calculation is referred to as annealing simulation [47-48].

In our calculations, five cycles were performed for each surface, testing 50,000 configurations at each cooling cycle. The temperature was controlled automatically. The geometry was optimized for each configuration, and the molecular positions varied according to pre-established probabilities using a Monte Carlo algorithm. Therefore, the probability of configuration m is given by the expression:

$$P_m = -C \, exp \left(\frac{-\beta E_m}{k_b T}\right)$$

where $C$ is a normalization constant, $\beta$ is the reciprocal temperature, $E_m$ is the total energy of configuration $m$, $k_b$ is the Boltzmann constant, and $T$ is the absolute temperature.

For the EB set, we defined equal probabilities (1/3) for the molecule to conform, rotate, and translate, aiming to identify the most favorable adsorption configurations on the cleavaged vivianites. The UFF force field, with the QEq method for charges, the atom-based method for van der Waals interactions, and the Ewald method for Coulombic interactions, were the same as those previously discussed for geometrically optimizing the systems. The adsorption space was considered the entire simulation box, and energy distributions were recorded, along with the ten energetically most favorable configurations.

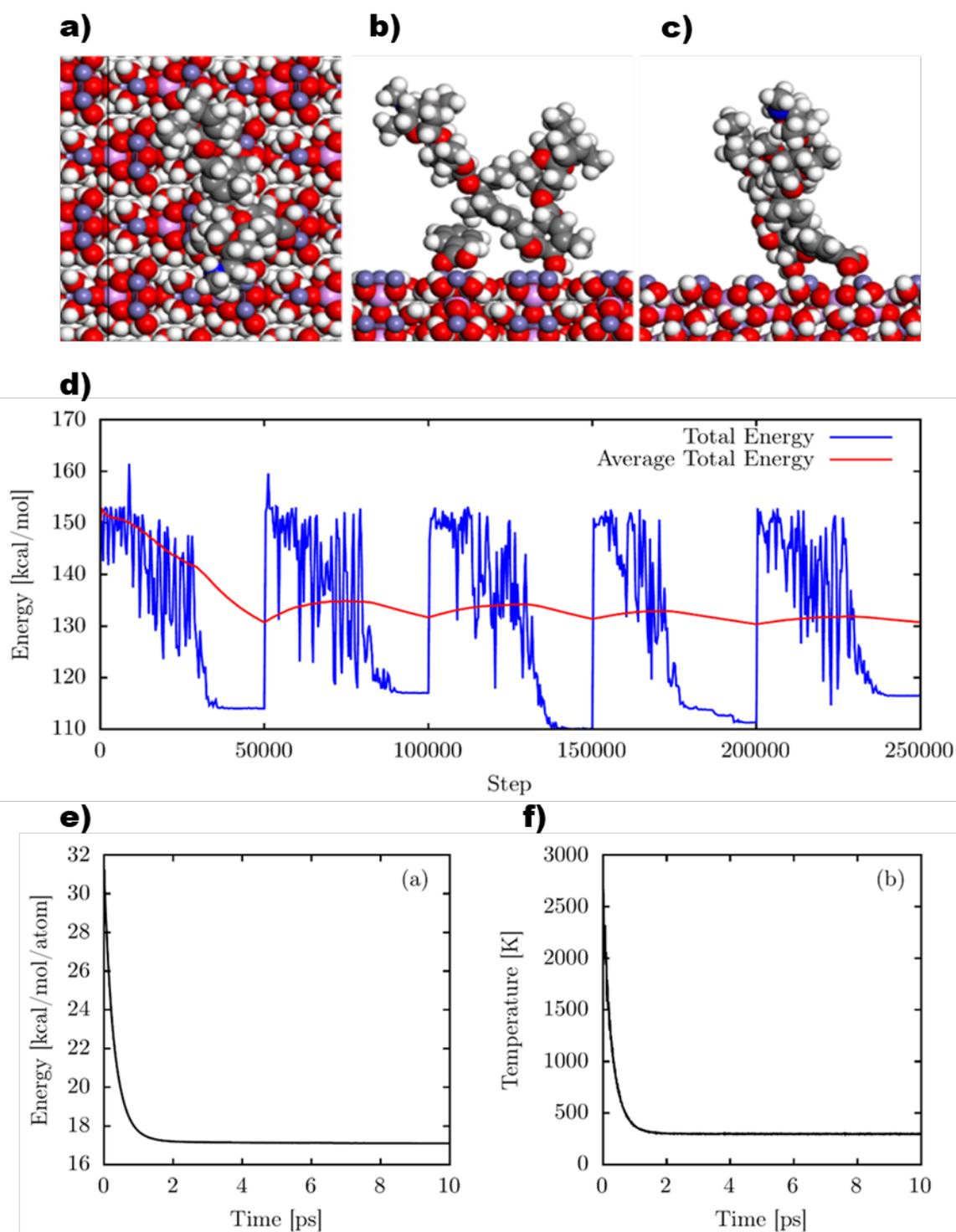

**Figure 5** : Front view (a) and side views (b, c) of EB adsorbed in the cleavaged vivianites, with a cleavage along the xy plane direction, for the configuration with the lowest energy among the 250,000 simulated configurations, (d) Total energy (blue) and cumulative average (red) for the annealing simulation of the adsorbate set (vivianite) and adsorbent (EB) after evaluating 250,000 configurations across five different temperature cycles, and (e) Total

energy (a) and temperature (b) as a function of simulation time for the xy plane of the cleavaged vivianites.

The most favorable adsorption configuration of EB on the xy direction of vivianites is presented in **Figure 5**, showing different views: a top view **(a)** and two side views **(b, c)**. The ten most favorable configurations obtained from the 250,000 calculated configurations exhibit similar characteristics, with both molecules positioned close to the substrate, except for stronger interactions in the region with oxygen atoms. The strongest interaction consistently occurs in the direction opposite to the nitrogen atom, indicating a vertical orientation of the insecticide relative to the vivianite in the xy plane. The calculated adsorption energy was -17.2 cal/mol/atom. **Figure 5d** illustrates the evolution of the annealing simulation for the EB configurations on the xy plane of the cleavaged vivianites. In this Figure, the temperature cycles are defined every 50,000 configurations (steps), demonstrating the method convergence in each cycle. The average total energy is approximately 130 kcal/mol.

For the xz direction, EB is adsorbed horizontally. However, the lowest energy configuration has the EB longitudinal axis perpendicular to the linear valleys in this cleavage (see **Figure S1b**), with an adsorption energy of -25.3 cal/mol/atom. Finally, when adsorbed along the yz direction, EB also adopts a vertical configuration, similar to the xy case, but with a smaller adsorption energy of -67.4 cal/mol/atom due to stronger interactions with P and Fe atoms, in contrast to the xy and xz cases, where the primary interactions are with H and O atoms.

Finally, to verify the structural dynamics stability of the system, we carried out MD simulations using the isothermal-isobaric ensemble (NPT), with a temperature of 298 K and external pressure in the plane directions set to zero. A time step of 1 fs was used, with a total simulation duration of 10,000 steps. A Berendsen thermostat maintained the temperature with a decay constant of 100 fs [49]. The Berendsen barostat was also used, with the same decay constant [50]. The force field, charge modeling, and non-bonded interactions were the same previously described. **Figure 5e** shows the convergence of energy (panel a) and temperature (panel b), with average values of 17.1 kcal/mol/atom and 298 K, respectively.

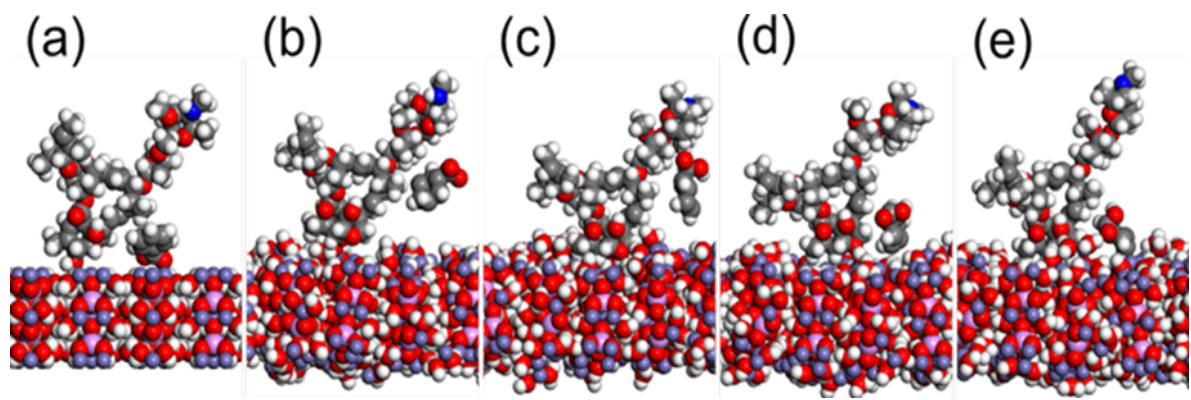

**Figure 6:** Representative MD snapshots of the EB adsorbed on the xy plane of the vivianite crystal at 0 ps (a), 2.5 ps (b), 5 ps (c), 7.5 ps (d), and 10 ps (e).

In **Figure 6,** we present representative MD snapshots at 0 ps (a), 2.5 ps (b), 5.0 ps (c), 7.5 ps (d), and 10 ps. Despite the expected temperature-induced atomic fluctuations, it is evident that EB continues interacting with the vivianites throughout the whole simulations, with no detachments from the substrate, and the vivianites also maintain their structural integrity. A comparative study with the previous detection method used for sensing EB are given in **Table 1.**

**Table 1** Comparison of the current study with previously employed analytical methods.

| Material Used | Method Used | Limit of detection | Reference |
|---|---|---|---|
| 1. formic acid (HCOOH) in water and 0.1% HCOOH in acetonitrile (CH3CN) | Liquid chromatography/Mass spectroscopy | 0.01 g/ml | [51] |
| 2. g-MWCNTs | Liquid chromatography/Mass spectroscopy | 50 µg/kg | [52] |
| 3. N-methylimidazole and Trifluoroacetic anhydride | Liquid chromatography-fluorescence detection | 0.001 mg/kg | [53] |
| 4. Vivianene/GCE | Spectroscopy and cyclic voltammetry | 10 pg/ml | This work |

**Summary and Conclusions**

In this study, we have utilized 2D vianenes, obtained from mechanically exfoliating of naturally occurring 3D vivianites (phosphate minerals), for the detection of emamectin benzoate (EB) molecules, an insecticide frequently applied in farming operations. The analyses were performed using spectroscopic and electrochemical detection techniques. The Raman technique demonstrated the ability to detect 50 µg/L of the analyte through a non-contact method, significantly below the allowable threshold. The FTIR analysis indicated the absence of new peaks, suggesting that the predominant interactions occurring are attributed to physical adsorption. This observation is further corroborated by the alteration in zeta potential, which shifted from 34 mV for vivianene to -4.7 mV upon the addition of EB to the solution. The electrochemical detection was performed using cyclic voltammetry. The anodic and cathodic peak currents in the case of CV were observed at approximately 0.41 V and -0.1 V, respectively, with a detection limit of 5 µg/L at lower scan rates. Fully atomistic molecular dynamics simulations were also used to gain further insights into the EB interaction mechanisms with the vivianites. The MD results corroborate the experimental data interpretation of an adsorption mechanism.

The application of natural 2D phosphate minerals offers significant advantages in terms of sensitivity, specificity, and tunability, making them highly effective for detecting insecticides at very low concentrations. The aforementioned qualities contribute significantly to the progress of advanced sensors aimed at enhancing agricultural monitoring and promoting sustainable development.

**References**

[1] F. Arduini, S. Cinti, V. Scognamiglio, D. Moscone, G. Palleschi, How cutting-edge technologies impact the design of electrochemical (bio)sensors for environmental analysis. A review, Anal Chim Acta 959 (2017) 15–42. https://doi.org/10.1016/j.aca.2016.12.035.

[2] X. Chen, M. Leishman, D. Bagnall, N. Nasiri, Nanostructured Gas Sensors: From Air Quality and Environmental Monitoring to Healthcare and Medical Applications, Nanomaterials 11 (2021) 1927. https://doi.org/10.3390/nano11081927.

[3] D. Tyagi, H. Wang, W. Huang, L. Hu, Y. Tang, Z. Guo, Z. Ouyang, H. Zhang, Recent advances in two-dimensional-material-based sensing technology toward health and environmental monitoring applications, Nanoscale 12 (2020) 3535–3559. https://doi.org/10.1039/C9NR10178K.

[4] P. Miró, M. Audiffred, T. Heine, An atlas of two-dimensional materials, Chem. Soc. Rev. 43 (2014) 6537–6554. https://doi.org/10.1039/C4CS00102H.

[5] C.N.R. Rao, A.K. Sood, K.S. Subrahmanyam, A. Govindaraj, Graphene: The New Two-Dimensional Nanomaterial, Angewandte Chemie International Edition 48 (2009) 7752–7777. https://doi.org/10.1002/anie.200901678.

[6] S. Varghese, S. Varghese, S. Swaminathan, K. Singh, V. Mittal, Two-Dimensional Materials for Sensing: Graphene and Beyond, Electronics (Basel) 4 (2015) 651–687. https://doi.org/10.3390/electronics4030651.

[7] C. Tan, X. Cao, X.-J. Wu, Q. He, J. Yang, X. Zhang, J. Chen, W. Zhao, S. Han, G.-H. Nam, M. Sindoro, H. Zhang, Recent Advances in Ultrathin Two-Dimensional Nanomaterials, Chem Rev 117 (2017) 6225–6331. https://doi.org/10.1021/acs.chemrev.6b00558.

[8] S. Slathia, C. Wei, M. Tripathi, R. Tromer, S.D. Negedu, C.S. Boland, S. Sarkar, D.S. Galvao, A. Dalton, C.S. Tiwary, Thickness dependent tribological and magnetic behavior of

two-dimensional cobalt telluride (CoTe$_2$), 2d Mater 11 (2024) 035006. https://doi.org/10.1088/2053-1583/ad3cec.

[9] E. Singh, M. Meyyappan, H.S. Nalwa, Flexible Graphene-Based Wearable Gas and Chemical Sensors, ACS Appl Mater Interfaces 9 (2017) 34544–34586. https://doi.org/10.1021/acsami.7b07063.

[10] I.J. Gómez, N. Alegret, A. Dominguez-Alfaro, M. Vázquez Sulleiro, Recent Advances on 2D Materials towards 3D Printing, Chemistry (Easton) 3 (2021) 1314–1343. https://doi.org/10.3390/chemistry3040095.

[11] X. Liu, T. Ma, N. Pinna, J. Zhang, Two-Dimensional Nanostructured Materials for Gas Sensing, Adv Funct Mater 27 (2017). https://doi.org/10.1002/adfm.201702168.

[12] D.J. Late, Y.-K. Huang, B. Liu, J. Acharya, S.N. Shirodkar, J. Luo, A. Yan, D. Charles, U. V. Waghmare, V.P. Dravid, C.N.R. Rao, Sensing Behavior of Atomically Thin-Layered MoS$_2$ Transistors, ACS Nano 7 (2013) 4879–4891. https://doi.org/10.1021/nn400026u.

[13] K.S. Novoselov, A.K. Geim, S. V. Morozov, D. Jiang, Y. Zhang, S. V. Dubonos, I. V. Grigorieva, A.A. Firsov, Electric Field Effect in Atomically Thin Carbon Films, Science (1979) 306 (2004) 666–669. https://doi.org/10.1126/science.1102896.

[14] R.L. Frost, M. Weier, Raman spectroscopic study of vivianites of different origins, Neues Jahrbuch Für Mineralogie - Monatshefte 2004 (2004) 445–463. https://doi.org/10.1127/0028-3649/2004/2004-0445.

[15] R. Frisenda, Y. Niu, P. Gant, M. Muñoz, A. Castellanos-Gomez, Naturally occurring van der Waals materials, NPJ 2D Mater Appl 4 (2020) 38. https://doi.org/10.1038/s41699-020-00172-2.

[16] T. ITO, Structure of Vivianite and Symplesite, Nature 164 (1949) 449–450. https://doi.org/10.1038/164449b0.

[17] World Health Organization and others, Pesticide residues in food-2013: toxicological evaluations, World Health Organization, 2015.

[18] D. Yang, B. Cui, C. Wang, X. Zhao, Z. Zeng, Y. Wang, C. Sun, G. Liu, H. Cui, Preparation and Characterization of Emamectin Benzoate Solid Nanodispersion, J Nanomater 2017 (2017) 1–9. https://doi.org/10.1155/2017/6560780.

[19] T. Yen, J. Lin, Acute Poisoning with Emamectin Benzoate, J Toxicol Clin Toxicol 42 (2004) 657–661. https://doi.org/10.1081/CLT-200026968.

[20] K.J. Willis, N. Ling, The toxicity of emamectin benzoate, an aquaculture pesticide, to planktonic marine copepods, Aquaculture 221 (2003) 289–297. https://doi.org/10.1016/S0044-8486(03)00066-8.

[21] J.W. Bloodworth, M.C. Baptie, K.F. Preedy, J. Best, Negative effects of the sea lice therapeutant emamectin benzoate at low concentrations on benthic communities around Scottish fish farms, Science of The Total Environment 669 (2019) 91–102. https://doi.org/10.1016/j.scitotenv.2019.02.430.

[22] D.S. Biovia, Material Studio 2020, Dassault Systemes, (2020).

[23] J. Fliege, B.F. Svaiter, Steepest descent methods for multicriteria optimization, Mathematical Methods of Operations Research (ZOR) 51 (2000) 479–494. https://doi.org/10.1007/s001860000043.

[24] A.K. Rappe, C.J. Casewit, K.S. Colwell, W.A. Goddard, W.M. Skiff, UFF, a full periodic table force field for molecular mechanics and molecular dynamics simulations, J Am Chem Soc 114 (1992) 10024–10035. https://doi.org/10.1021/ja00051a040.

[25] A.K. Rappe, W.A. Goddard, Charge equilibration for molecular dynamics simulations, J Phys Chem 95 (1991) 3358–3363. https://doi.org/10.1021/j100161a070.

[26] P.P. Ewald, Die Berechnung optischer und elektrostatischer Gitterpotentiale, Ann Phys 369 (1921) 253–287. https://doi.org/10.1002/andp.19213690304.

[27] U. Essmann, L. Perera, M.L. Berkowitz, T. Darden, H. Lee, L.G. Pedersen, A smooth particle mesh Ewald method, J Chem Phys 103 (1995) 8577–8593. https://doi.org/10.1063/1.470117.

[28] X. Wu, Z. Liu, W. Zhu, External electric field induced conformational changes as a buffer to increase the stability of CL-20/HMX cocrystal and its pure components, Mater Today Commun 26 (2021) 101696. https://doi.org/10.1016/j.mtcomm.2020.101696.

[29] S.L. Waddy, V.A. Merritt, M.N. Hamilton-Gibson, D.E. Aiken, L.E. Burridge, Relationship between dose of emamectin benzoate and molting response of ovigerous American lobsters (Homarus americanus), Ecotoxicol Environ Saf 67 (2007) 95–99. https://doi.org/10.1016/j.ecoenv.2006.05.002.

[30] B.R. McCammon CA, The oxidation mechanism of vivianite as studies by Möessbauer spectroscopy., American Mineralogist, 1980.

[31] Y. Wu, J. Luo, Q. Zhang, M. Aleem, F. Fang, Z. Xue, J. Cao, Potentials and challenges of phosphorus recovery as vivianite from wastewater: A review, Chemosphere 226 (2019) 246–258. https://doi.org/10.1016/j.chemosphere.2019.03.138.

[32] B. Piriou, J.F. Poullen, Raman study of vivianite, Journal of Raman Spectroscopy 15 (1984) 343–346. https://doi.org/10.1002/jrs.1250150510.

[33] H. Dai, W. Xu, Y. Chen, M. Li, Z. Chen, B. Yang, S. Mei, W. Zhang, F. Xie, W. Wei, R. Guo, G. Zhang, Narrow band-gap cathode $Fe_3(PO_4)_2$ for sodium-ion battery with enhanced sodium storage, Colloids Surf A Physicochem Eng Asp 591 (2020) 124561. https://doi.org/10.1016/j.colsurfa.2020.124561.

[34] Y. Wang, D.J. Asunskis, P.M.A. Sherwood, Iron (II) Phosphate ($Fe_3(PO_4)_2$ by XPS, Surface Science Spectra 9 (2002) 91–98. https://doi.org/10.1116/11.20030105.

[35] K. R., A.K. Singh, S. Das, S. Sarkar, T.K. Kundu, S. Kar, P.R. Sreeram, C.S. Tiwary, Giant Stark effect assisted radio frequency energy harvesting using atomically thin earth-abundant

iron sulphide (FeS$_2$), J Mater Chem A Mater 12 (2024) 8940–8951. https://doi.org/10.1039/D3TA07906F.

[36] B. Zhang, L. Wang, Y. Li, Fractionation and identification of iron-phosphorus compounds in sewage sludge, Chemosphere 223 (2019) 250–256. https://doi.org/10.1016/j.chemosphere.2019.02.052.

[37] M. Wu, J. Liu, B. Gao, M. Sillanpää, Phosphate substances transformation and vivianite formation in P-Fe containing sludge during the transition process of aerobic and anaerobic conditions, Bioresour Technol 319 (2021) 124259. https://doi.org/10.1016/j.biortech.2020.124259.

[38] M. Sang, J. Weng, X. Chen, G. Nie, Renewable cellulose aerogel embedded with nano-HFO for preferable phosphate capture from aqueous solution, Environmental Science and Pollution Research 30 (2022) 26613–26624. https://doi.org/10.1007/s11356-022-24087-1.

[39] E. Schütze, S. Gypser, D. Freese, Kinetics of Phosphorus Release from Vivianite, Hydroxyapatite, and Bone Char Influenced by Organic and Inorganic Compounds, Soil Syst 4 (2020) 15. https://doi.org/10.3390/soilsystems4010015.

[40] P.V. Campos, A.R.L. Albuquerque, R.S. Angélica, S.P.A. Paz, FTIR spectral signatures of amazon inorganic phosphates: Igneous, weathering, and biogenetic origin, Spectrochim Acta A Mol Biomol Spectrosc 251 (2021) 119476. https://doi.org/10.1016/j.saa.2021.119476.

[41] A. Elabasy, A. Shoaib, M. Waqas, M. Jiang, Z. Shi, Synthesis, Characterization, and Pesticidal Activity of Emamectin Benzoate Nanoformulations against Phenacoccus solenopsis Tinsley (Hemiptera: Pseudococcidae), Molecules 24 (2019) 2801. https://doi.org/10.3390/molecules24152801.

[42] Y. Long, H. Zhang, G. Liao, M. Chen, X. Chen, L. Qin, C. Chen, Z. Chen, X. Wu, F. Zhu, Distribution of Emamectin Benzoate Granules in Maize Plants by Broadcasting into Maize Leaf Whorls, ACS Omega 8 (2023) 4209–4219. https://doi.org/10.1021/acsomega.2c07402.

[43] Y. Wang, A. Wang, C. Wang, B. Cui, C. Sun, X. Zhao, Z. Zeng, Y. Shen, F. Gao, G. Liu, H. Cui, Synthesis and characterization of emamectin-benzoate slow-release microspheres with different surfactants, Sci Rep 7 (2017) 12761. https://doi.org/10.1038/s41598-017-12724-6.

[44] F. Zhu, H. Zhang, C. Chen, Y. Long, G. Liao, M. Chen, L. Qin, X. Chen, Y. He, Z. Chen, Controlled-release alginate-bentonite polymer gel granules of emamectin benzoate and control efficacy against *Spodoptera frugiperda*, Pest Manag Sci 79 (2023) 324–335. https://doi.org/10.1002/ps.7202.

[45] S.-H. Choi, H.G. Park, Surface-enhanced Raman scattering (SERS) spectra of sodium benzoate and 4-picoline in Ag colloids prepared by γ-irradiation, Appl Surf Sci 243 (2005) 76–81. https://doi.org/10.1016/j.apsusc.2004.09.051.

[46] Z. Cao, C. Wang, Y. Sun, M. Liu, W. Li, J. Zhang, Y. Fu, A Ru/RuO$_2$ heterostructure boosting electrochemistry-assisted selective benzoic acid hydrogenation, Chem Sci 15 (2024) 1384–1392. https://doi.org/10.1039/D3SC05312A.

[47] R.L.C. Akkermans, N.A. Spenley, S.H. Robertson, Monte Carlo methods in Materials Studio, Mol Simul 39 (2013) 1153–1164. https://doi.org/10.1080/08927022.2013.843775.

[48] H. Soleimani, M.K. Baig, N. Yahya, L. Khodapanah, M. Sabet, B.M.R. Demiral, M. Burda, Synthesis of ZnO nanoparticles for oil–water interfacial tension reduction in enhanced oil recovery, Applied Physics A 124 (2018) 128. https://doi.org/10.1007/s00339-017-1510-4.

[49] A.S. Lemak, N.K. Balabaev, On The Berendsen Thermostat, Mol Simul 13 (1994) 177–187. https://doi.org/10.1080/08927029408021981.

[50] Y. Lin, D. Pan, J. Li, L. Zhang, X. Shao, Application of Berendsen barostat in dissipative particle dynamics for nonequilibrium dynamic simulation, J Chem Phys 146 (2017). https://doi.org/10.1063/1.4978807.

[51] R.S. Raj, S.V. Krishnamoorthy, A. Suganthi, C. Kavitha, S. Karthikeyan, A.K. Karedla, P. Karthik, J. Kousika, Method validation and monitoring of emamectin benzoate in mature

banana fruit with peel and pulp through Liquid chromatography-Mass spectrometry/ Mass spectrometry (LC-MS/MS), Journal of Applied and Natural Science 15 (2023) 1230–1236. https://doi.org/10.31018/jans.v15i3.4828.

[52] Z. Xu, L. Li, Y. Xu, S. Wang, X. Zhang, T. Tang, J. Yu, H. Zhao, S. Wu, C. Zhang, X. Zhao, Pesticide multi-residues in Dendrobium officinale Kimura et Migo: Method validation, residue levels and dietary exposure risk assessment, Food Chem 343 (2021) 128490. https://doi.org/10.1016/j.foodchem.2020.128490.

[53] Y. Liu, H. Sun, X. Wang, H. Chang, S. Wang, Dissipation Dynamic, Residue Distribution and Risk Assessment of Emamectin Benzoate in Longan by High-Performance Liquid Chromatography with Fluorescence Detection, Molecules 28 (2023) 3346. https://doi.org/10.3390/molecules28083346.